\documentclass[a4paper,10pt]{article}

\usepackage{publication}

\usepackage[T1]{fontenc}
\usepackage[utf8]{inputenc}
\usepackage{lmodern}
\usepackage{microtype}
\usepackage{booktabs}
\usepackage{longtable}
\usepackage{tabularx}
\usepackage{array}
\usepackage{enumitem}
\usepackage{listings}
\usepackage{xcolor}
\usepackage{caption}
\usepackage{amsmath}
\usepackage{csquotes}
\usepackage[backend=biber,style=numeric,sorting=nyt,maxbibnames=99]{biblatex}
\setlist[itemize]{leftmargin=*,nosep}
\setlist[enumerate]{leftmargin=*,nosep}
\lstdefinestyle{jsonstyle}{
  basicstyle=\ttfamily\footnotesize,
  breaklines=true,
  columns=fullflexible,
  frame=single,
  showstringspaces=false
}
\title{GCVE: A Decentralized Model for Vulnerability Identification, Publication, and Operational Enrichment}
\author[1]{Alexandre Dulaunoy}
\contact{alexandre.dulaunoy@circl.lu}
\affil[1]{GCVE initiative and Computer Incident Response Center Luxembourg (CIRCL)}
\date{Preprint version of \today}

\begin{document}
\maketitle

\noindent\textbf{Preprint note.} This manuscript is a preprint prepared for Pass the SALT 2026 in Lille, France, accompanying the conference session \enquote{GCVE: Rebooting Vulnerability Tracking for an Open Security Ecosystem}. The session is listed in the Pass the SALT 2026 programme for 2 July 2026 at 11:10--11:45 in Amphitheater 122, and the 2026 edition of the conference is held at Universit\'e Catholique de Lille in Lille, France from 30 June to 2 July 2026 \cite{pts2026-gcve-talk,pts2026-conference}. This preprint is not a formal proceedings publication; it is intended as a research-grade companion paper for discussion, implementation review, and further standardization work.

\begin{abstract}
The Global CVE (GCVE) initiative defines a decentralized model for vulnerability identification and publication in which GCVE Numbering Authorities (GNAs) allocate identifiers and publish records autonomously while remaining interoperable through shared Best Current Practices (BCPs). This paper describes the design and implementation of GCVE from its initial identifier-allocation motivation to its current state as a full-featured vulnerability publication ecosystem. We analyze the GCVE identifier model, the GNA autonomy model, the signed directory, decentralized publishing, the practical vulnerability-handling guidance, the GCVE record container, the BCP extension mechanism for AI-assisted processing, automatically enriched data streams, distributed Known Exploited Vulnerability (KEV) assertions, and the reference implementation in vulnerability-lookup. The central argument is that GCVE shifts vulnerability infrastructure from a single canonical pipeline toward a federated network of independently governed, machine-readable assertions. This design supports diverse use cases--including original vulnerability reports, downstream product impact statements, remediation records, clarification records, service-specific exposure statements, KEV assertions, and machine-generated enrichment--without requiring all participants to adopt the same disclosure ideology or operational workflow.
\end{abstract}

\section{Introduction}

The vulnerability information ecosystem is not a single uniform system. One part of it is organized around centralized or centrally coordinated registries in which identifier allocation, record publication, and quality control are governed through explicit program rules. The CVE Program, for example, identifies, defines, and catalogs publicly disclosed cybersecurity vulnerabilities, while CVE Numbering Authorities operate within defined scopes and are subject to operational rules intended to preserve consistency across the shared namespace \cite{cve-home,cve-cnas,cve-cna-rules}. This model remains essential: rigorous control helps create stable identifiers, familiar workflows, and widely reused records. At the same time, central coordination can impose workflow assumptions, eligibility boundaries, review expectations, and publication paths that do not fit every disclosure, downstream impact statement, product-specific clarification, or operational exploitation signal.

A second part of the ecosystem is advisory-first and publisher-driven. Vendors, open-source projects, researchers, security teams, Linux distributions, national CSIRTs, cloud providers, and community groups often publish vulnerability information through their own advisories, repositories, mailing lists, websites, security pages, or machine-readable feeds. These publications may be accurate and operationally important, but they are frequently isolated from a shared discovery and correlation layer. They can be difficult to ingest at scale, hard to relate to other identifiers, and invisible to consumers that only follow a single registry. This is the operational gap that vulnerability-lookup addresses: it gathers multiple vulnerability sources, correlates records independently of any single identifier scheme, and exposes an API, feeds, sightings, comments, bundles, and GCVE support as practical infrastructure for multi-source vulnerability intelligence \cite{vl-home,vl-github}.

GCVE was announced on 16 April 2025 as a decentralized approach to identifying and numbering security vulnerabilities. Its initial objective was pragmatic: allow independent GCVE Numbering Authorities (GNAs) to assign vulnerability identifiers directly, improve allocation speed and autonomy, and remain compatible with the existing CVE ecosystem by reserving GNA ID 0 for standard CVEs \cite{gcve-announce-2025,gcve-about}. That initial allocation idea has since become a broader publication model. The current GCVE ecosystem includes a signed GNA directory, a decentralized publication standard, an extensible vulnerability record container, a GNA conformance model, a distributed KEV assertion format, AI-assisted provenance metadata, enriched dumps, improved CPE modeling, a public database instance, and the vulnerability-lookup reference implementation \cite{gcve-bcp-index,gcve-db-announce,gcve-recent-activities-2026}.

As a companion preprint to the Pass the SALT 2026 presentation on GCVE, this paper makes five contributions. First, it describes the design principles behind GCVE: autonomy, compatibility, decentralization, transparency, extensibility, and operational reimplementability. Second, it summarizes all current GCVE BCPs and explains how they fit together. Third, it explains why practical vulnerability-handling guidance is part of the model rather than an external afterthought. Fourth, it gives special attention to GCVE-BCP-09, which explicitly broadens the scope of a GCVE record beyond a traditional vulnerability description. Fifth, it explains how vulnerability-lookup operationalizes GCVE by verifying the directory, gathering decentralized publications, publishing local records, exposing KEV catalogs, integrating CISA and ENISA KEV sources, supporting CNA/GNA-compatible workflows, and enabling enriched data streams \cite{gcve-bcp-02,vl-gcve-user-manual,vl-3-0-0,vl-5-0-0}.

\section{Motivation and Origins}

The GCVE initiative emerged from operational experience with vulnerability collection, correlation, publication, and standardization. The problem is not merely the availability of an identifier string. It is the end-to-end ability for a publisher to assign an identifier, publish structured data, establish relationships with other records, add evidence and metadata, and allow consumers to decide which publishers and signals they trust. GCVE therefore treats vulnerability publication as an ecosystem problem: identifiers, records, relationships, discovery metadata, trust signals, enrichment, and reference tooling must evolve together.

The design also reflects lessons learned from the MISP project and the later MISP standardization effort. The original MISP work showed that a practical, open-source implementation can anchor a collaborative information-sharing model while still allowing communities to keep local control over distribution and trust \cite{wagner2016misp}. The MISP format then evolved into an open, practical JSON-based standard, explicitly built from operational use cases and implementation experience so that other tools could interoperate without being forced to run the same software stack \cite{misp-datamodels,misp-standard}. GCVE follows the same engineering philosophy for vulnerability information: begin from operational needs, document the interoperable parts as open BCPs, keep the format implementable by independent tools, and preserve room for communities to extend the model when new use cases appear.

Three constraints shaped the design.

\begin{enumerate}
  \item \textbf{Compatibility with existing CVE practices.} GCVE is designed to complement CVE, not replace it. Standard CVEs can be represented through reserved GNA ID 0, allowing GCVE-aware tooling to correlate legacy and new identifiers \cite{gcve-about,gcve-bcp-04}.
  \item \textbf{Operational autonomy.} GNAs should be able to allocate identifiers at their own pace, define internal policies, and publish records without waiting for a centralized allocation or publication bottleneck \cite{gcve-about,gcve-bcp-06}.
  \item \textbf{Machine-readable federation.} Autonomy requires enough structure for aggregation, validation, indexing, and selective trust. GCVE therefore defines BCPs for directory verification, publication endpoints, record format, GNA conformance, and relationship semantics \cite{gcve-bcp-index,gcve-bcp-01,gcve-bcp-03,gcve-bcp-05,gcve-bcp-06}.
\end{enumerate}

The result is a shift from a centralized vulnerability database mindset to a distributed assertion network. In that network, a vulnerability identifier is not the only artifact. A GNA can publish an original advisory, a correction, a vendor-specific impact statement, a remediation object, a translation, a detection statement, a KEV assertion, or an enriched record. Consumers can then correlate these objects by identifiers, relationships, source, record type, evidence, and trust policy.

\section{Design Principles}

GCVE is built around six principles.

\subsection{Decentralized Allocation}

Each GNA receives a numeric identifier. The recommended GCVE identifier form is \texttt{GCVE-<GNA-ID>-<YEAR>-<UNIQUE-ID>}, while alternative GNA-specific values are also permitted as long as they preserve the \texttt{GCVE-<GNA-ID>-<GNA-VALUE>} prefix structure and use valid characters \cite{gcve-about,gcve-bcp-04}. This design separates global uniqueness from central assignment. The globally meaningful portion is the GNA namespace; the local allocation policy is delegated to the GNA.

\subsection{Compatibility Without Subordination}

GCVE preserves compatibility with CVE by reserving GNA ID 0 for existing CVE identifiers \cite{gcve-about,gcve-announce-2025}. Compatibility is important because vulnerability data consumers already operate large CVE-based workflows. However, compatibility is not subordination: a GCVE record can be assigned for cases that CVE-style records do not model well, including clarifications, downstream statements, remediation-focused records, or service-specific exposure statements \cite{gcve-bcp-09}.

\subsection{Transparency Over Uniformity}

GCVE-BCP-06 captures a core governance principle: the ecosystem should not impose a single disclosure ideology or review model. Instead, it requires GNAs to document their operational and visibility models so that consumers can make informed decisions \cite{gcve-bcp-06}. The practical result is a system in which an automated allocator, a vendor PSIRT, a research publisher, a community reviewer, or a private-public hybrid publisher can coexist, provided that each is transparent about scope, process, and publication behavior.

\subsection{Trust as a Consumer Policy}

GCVE does not require every consumer to trust every GNA equally. BCP-03 explicitly places publication under GNA control and discovery under the directory, while consumers select the sources they wish to pull from \cite{gcve-bcp-03}. This turns trust from a universal property into a local policy decision. Vulnerability-lookup implements this by allowing instances to select GNAs and feeds, aggregate selected sources, and correlate records locally \cite{gcve-bcp-03,vl-home,vl-gcve-user-manual}.

\subsection{Extensibility by Ignorable Additions}

The GCVE record format is derived from the CVE Record Format, but adds GCVE-specific fields and extension points such as \texttt{x\_gcve} and \texttt{x\_vulnerability-lookup}. Unknown record types and extension keys are intended to be safely ignored by consumers that do not understand them \cite{gcve-bcp-05,gcve-bcp-05-x-01,gcve-enriched-dumps}. This makes schema evolution possible without forcing lockstep upgrades across the ecosystem.

\subsection{Reimplementability}

GCVE standards deliberately rely on common mechanisms: JSON records, HTTP REST APIs, static files, detached signatures, and public keys \cite{gcve-bcp-01,gcve-bcp-03}. This choice keeps the barrier to entry low. A minimal GNA can publish a static dump; a larger GNA can publish an API; an aggregator can use the same directory fields to discover both.

\section{The GCVE BCP Corpus}

The GCVE BCPs are community-driven guidelines that document recommended procedures, configurations, and operational principles for secure, reliable, consistent GCVE infrastructure. They are descriptive rather than prescriptive, and adherence is strongly recommended but not mandatory \cite{gcve-bcp-index}. Table~\ref{tab:bcps} summarizes the current BCP corpus as of 30 May 2026. The index currently lists BCP-01 through BCP-07, BCP-09, BCP-10, and extension BCP-05-X-01; no BCP-08 is listed in the public index at the time of this draft \cite{gcve-bcp-index}.

\begin{longtable}{p{0.12\linewidth}p{0.31\linewidth}p{0.49\linewidth}}
\caption{Current GCVE BCP corpus and architectural role.}\label{tab:bcps}\\
\toprule
\textbf{BCP} & \textbf{Title} & \textbf{Role in the GCVE architecture}\\
\midrule
\endfirsthead
\toprule
\textbf{BCP} & \textbf{Title} & \textbf{Role in the GCVE architecture}\\
\midrule
\endhead
BCP-01 & Signature Verification of the Directory File & Defines integrity and authenticity verification for the GNA directory using a detached SHA-512 signature and public key distribution. It anchors directory trust without turning publication itself into a centralized workflow \cite{gcve-bcp-01}.\\
BCP-02 & Practical Guide to Vulnerability Handling and Disclosure & Provides operational guidance for vulnerability intake, coordination, remediation, and communication. It connects GCVE to practical coordinated vulnerability disclosure workflows \cite{gcve-bcp-02}.\\
BCP-03 & Decentralized Publication Standard & Defines how GNAs publish directly via REST APIs or static files, and how consumers discover pull endpoints from the directory. It is the main federation mechanism \cite{gcve-bcp-03}.\\
BCP-04 & Recommendations and Best Practices for ID Allocation & Defines recommended GCVE ID syntax, alternative GNA-specific forms, length guidance, and use of GCVE IDs for complementary and relationship-based information \cite{gcve-bcp-04}.\\
BCP-05 & GCVE Vulnerability Format & Defines the GCVE vulnerability container, derived from the CVE JSON record format, and introduces GCVE-specific extension and relationship mechanisms \cite{gcve-bcp-05}.\\
BCP-06 & GNA Requirements and Evaluation Criteria & Defines transparent operational expectations for GNAs, including disclosure model, visibility, synchronization endpoint availability, identifier stability, and conformance metadata \cite{gcve-bcp-06}.\\
BCP-07 & Known Exploited Vulnerability Assertion Format & Defines KEV as an attributable, event-level assertion separate from vulnerability identity, supporting multiple sources, statuses, evidence, and confidence \cite{gcve-bcp-07}.\\
BCP-09 & Scope of a GCVE Record & Clarifies that a GCVE record is any GNA-assigned vulnerability-related information object, not only a classical vulnerability description \cite{gcve-bcp-09}.\\
BCP-10 & Improved Common Platform Enumeration for GCVE & Defines an improved platform enumeration model aligned with cpe-editor, using stable UUIDs, deterministic imports, aliases, relationships, provenance, and moderation \cite{gcve-bcp-10}.\\
BCP-05-X-01 & AI-Assisted Vulnerability Information Annotation & Extends BCP-05 with metadata for AI-assisted or automated creation, enrichment, classification, and analysis, including review state and model provenance \cite{gcve-bcp-05-x-01}.\\
\bottomrule
\end{longtable}

\section{Practical Guidance and Open Socio-Technical Standards}

GCVE intentionally includes practice in the standard model. BCP-02, the \emph{Practical Guide to Vulnerability Handling and Disclosure}, is not merely a narrative companion to the identifier and publication specifications; it is part of the same architecture of autonomy and interoperability. The guide describes the operational life cycle around vulnerability work, including preparation, report intake, investigation, remediation, communication with reporters and users, coordinated vulnerability disclosure, advisory publication, and continuous improvement \cite{gcve-bcp-02}. These practices determine whether identifiers and records are useful in real disclosure work: a record without a reporting channel, triage process, remediation path, or communication practice can be technically valid while remaining operationally ineffective.

This placement is also a deliberate open-standard choice. Vulnerability disclosure and vulnerability handling are already described in standards such as ISO/IEC 29147 and ISO/IEC 30111, which respectively cover disclosure and handling processes for vendors \cite{iso29147,iso30111}. Those standards are useful reference points, but they are not the same kind of artifact as an openly amendable operational BCP. GCVE-BCP-02 is published as part of the GCVE BCP corpus under a CC-BY-4.0 license, making it directly accessible, redistributable, reviewable, and easier for practitioners to amend through the same public process as the technical GCVE specifications \cite{gcve-bcp-02,gcve-bcp-index}. In this sense, BCP-02 reduces the distance between formal process knowledge and the communities that must implement it: open-source maintainers, smaller vendors, researchers, CSIRTs, GNAs, and public-interest infrastructure operators.

The socio-technical dimension should not be underestimated. GCVE treats vulnerability publication as a set of tools that should increase local agency rather than force every actor into a single institutional dependency. This orientation is close to Ivan Illich's idea of convivial tools: tools should expand the user's capacity to act autonomously, creatively, and responsibly rather than concentrating expertise and control in closed institutions \cite{illich1973tools}. Applied to vulnerability coordination, this means that a GNA should be able to allocate, publish, correct, enrich, and contextualize records with tools it can understand and operate, while consumers remain free to decide which sources they trust. BCP-02 provides the practical vocabulary for this autonomy: it connects identifiers, publication endpoints, and records to human practices of acknowledgement, coordination, remediation, safe communication, and public advisory writing.

This is why practical guidance belongs beside the data formats. BCP-03 and BCP-05 make records publishable and machine-readable; BCP-06 makes GNA operational models explicit; BCP-09 broadens the semantics of what may be published; and BCP-02 explains how vulnerability handling can be performed as a transparent, responsible, and improvable process. The GCVE model therefore standardizes not only objects, but also the minimum shared practices that allow independently governed actors to cooperate without surrendering operational autonomy.

\section{Core Architecture}

\begin{figure}[ht]
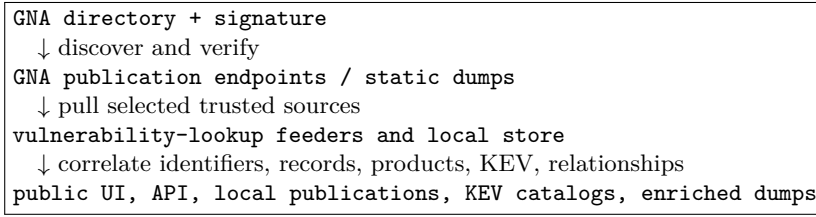

\centering
\fbox{%
\begin{minipage}{0.92\linewidth}
\small
\texttt{GNA directory + signature}\\
\hspace*{1em}$\downarrow$ discover and verify\\
\texttt{GNA publication endpoints / static dumps}\\
\hspace*{1em}$\downarrow$ pull selected trusted sources\\
\texttt{vulnerability-lookup feeders and local store}\\
\hspace*{1em}$\downarrow$ correlate identifiers, records, products, KEV, relationships\\
\texttt{public UI, API, local publications, KEV catalogs, enriched dumps}
\end{minipage}}
\caption{Simplified GCVE data flow. The directory discovers GNA endpoints; consumers decide which publishers to trust; vulnerability-lookup operationalizes collection, publication, correlation, and derived feeds.}
\label{fig:architecture}
\end{figure}

Figure~\ref{fig:architecture} summarizes the GCVE architecture. The signed directory is a discovery and trust anchor; GNA publication endpoints and dumps are the decentralized data plane; BCP-05 records and BCP-07 KEV assertions are the principal exchange objects; vulnerability-lookup is a reference implementation and operational aggregator; enriched dumps are a reproducible downstream data product.

\subsection{Signed GNA Directory}

The GNA directory contains registered GNA metadata such as GNA ID, short name, full name, publication URLs, API endpoints, dump endpoints, allocation process links, and pull API locations \cite{gcve-about,gcve-bcp-01,gcve-bcp-03}. BCP-01 defines verification of the directory file. The directory is published as JSON and accompanied by a detached SHA-512 signature, with public keys available over HTTP and DNS \cite{gcve-bcp-01}. The design provides a lightweight authenticity mechanism for the directory while leaving vulnerability publication to each GNA.

\subsection{Identifier Namespace}

The identifier model balances global uniqueness and GNA autonomy. The GNA ID creates a globally unique namespace, while the GNA controls local allocation. BCP-04 recommends \texttt{GCVE-<GNA-ID>-<YEAR>-<UNIQUE-ID>} but permits alternative local values to preserve operational flexibility \cite{gcve-bcp-04}. Importantly, BCP-04 also recognizes that multiple identifiers can describe complementary aspects of the same underlying vulnerability, including patches, remediation, parent-child relationships, or alternative perspectives \cite{gcve-bcp-04}. This design avoids forcing all vulnerability-related information into a single canonical record.

\subsection{Publication Plane}

BCP-03 defines the decentralized publication plane. A GNA can publish directly without a central database, using REST APIs or static files. The GCVE directory advertises the relevant pull endpoint, and clients pull from the GNAs they trust \cite{gcve-bcp-03}. The BCP-03 model is intentionally minimal: it specifies a common endpoint such as \texttt{/api/gcve/publication}, optional filters, and machine-readable output. The same model supports small publishers with static dumps and large publishers with APIs.

\subsection{Record Plane}

BCP-05 defines the GCVE vulnerability format. The design derives from CVE JSON v5 to preserve familiarity and compatibility, while introducing GCVE-specific objects and extensions \cite{gcve-bcp-05}. Record types can include advisories, updates, analyses, metadata, references, comments, statements, remediations, deprecations, detections, translations, and bundles; unknown types are to be ignored by consumers for forward compatibility \cite{gcve-bcp-05}. This record layer is the foundation for the more expansive scope formalized by BCP-09.

\section{GNA Autonomy Model}

The core innovation of GCVE is not merely a new identifier prefix. It is a governance and publication model in which autonomy is explicit, documented, and machine-actionable.

\subsection{Operational Diversity}

BCP-06 recognizes that GNAs differ in disclosure philosophy, review process, organization structure, resources, and legal constraints \cite{gcve-bcp-06}. Rather than forcing a single workflow, it asks GNAs to declare their operational model and visibility model. Examples include automated allocator, community reviewer, vendor authority, research publisher, immediate full disclosure, and private-public publication models \cite{gcve-bcp-06}. The system therefore admits both high-touch coordinated disclosure and rapid automated publication, provided that the publisher is transparent about what it does.

\subsection{Autonomy with Accountability}

Autonomy is bounded by transparency. BCP-06 requires public conformance information, stable contact points, declared scope, publication visibility, synchronization endpoint availability, identifier stability, and machine-readable metadata \cite{gcve-bcp-06}. It explicitly emphasizes that conformance should evaluate stability, transparency, interoperability, and operational reliability rather than judge the disclosure ideology itself \cite{gcve-bcp-06}. This is a crucial distinction. GCVE does not seek one global vulnerability policy; it seeks a network of clear, interoperable policies.

\subsection{Consumer-Side Trust}

Because GNAs differ, consumers need local trust policy. A national CSIRT may pull one set of GNAs; an enterprise may select vendors and KEV catalogs relevant to its assets; a researcher may ingest all public sources and then filter analytically. The GCVE directory and BCP-03 publication model support this by making publisher metadata and endpoints discoverable, while vulnerability-lookup provides a reference implementation for selective collection and correlation \cite{gcve-bcp-03,vl-home,vl-gcve-user-manual}.

\section{The Scope of a GCVE Record}

GCVE-BCP-09 is essential to understanding the project. It states that a GCVE record is not limited to a traditional vulnerability description; it is a GNA-assigned vulnerability-related information object \cite{gcve-bcp-09}. This definition turns GCVE into a standard for publishing vulnerability knowledge, not merely assigning vulnerability IDs.

\subsection{Why the Traditional One-ID/One-Description Model Is Too Narrow}

The traditional assumption that one identifier maps to one canonical vulnerability description is often insufficient. Modern vulnerability management must handle cloud services, managed services, downstream products, configuration-specific exposure, remediation-only guidance, delayed vendor statements, duplicate or conflicting reports, exploitation status, and rapidly changing operational context. BCP-09 identifies these cases and clarifies that the record meaning comes from the assigning GNA, the GNA's published model, the record content, and explicit relationships \cite{gcve-bcp-09}.

\subsection{Record Use Cases}

BCP-09 permits a wide range of vulnerability-related records, including the following:

\begin{itemize}
  \item \textbf{Classical vulnerability records} for software, firmware, hardware, protocols, or infrastructure.
  \item \textbf{Cloud and managed-service records} where the affected object may not be a shipped product version but a service, tenant configuration, API, or managed deployment.
  \item \textbf{Clarification records} that refine, correct, or contextualize another record without claiming to replace the original authority.
  \item \textbf{Remediation or mitigation records} that assign durable identifiers to patches, configuration changes, workarounds, or operational controls.
  \item \textbf{Product-, vendor-, or deployment-specific impact statements} that explain whether a known issue affects a downstream distribution, appliance, service, or deployment pattern.
  \item \textbf{Relationship and context records} that express equivalence, dependency, containment, parent-child relationships, supersession, or deprecation.
  \item \textbf{Other GNA-defined vulnerability information} when the GNA documents its model and the record remains within a vulnerability-related scope.
\end{itemize}

This design embraces pluralism. A GCVE record can be an original claim, a downstream statement, an operational observation, or a structured enrichment. Consumers should not assume that all records have identical semantics; they should interpret a record in its GNA context and follow explicit relationships \cite{gcve-bcp-09}.

\subsection{Relationship Semantics}

A broad record scope requires explicit relationships. Otherwise, consumers may mistake a clarification for a duplicate, a remediation object for a vulnerability, or a product-specific non-affected statement for a universal claim. BCP-05 and BCP-09 together encourage relationship fields and typed records so that tools can build a graph of vulnerability knowledge rather than a flat table of descriptions \cite{gcve-bcp-05,gcve-bcp-09}. This is also why vulnerability-lookup's ability to correlate across databases and identifiers is central to the reference implementation \cite{gcve-db-announce,vl-gcve-user-manual}.

\section{Extensibility and AI-Assisted Metadata}

\subsection{BCP Extension Mechanism}

BCP-05-X-01 demonstrates how GCVE extends the record format without breaking consumers. The extension defines metadata for AI-assisted or automated processing in record creation, enrichment, classification, transformation, and analysis \cite{gcve-bcp-05-x-01}. It can be attached at record level or field level, typically under \texttt{x\_gcve[].extensions["bcp-05-x-01"]}. Key fields include scope, GNA source, field name, tags, description, AI level, review status, and model metadata \cite{gcve-bcp-05-x-01}.

The key design choice is provenance, not automation for its own sake. AI-assisted vulnerability data can be useful, but consumers need to know where it came from, which model produced it, whether humans reviewed it, and which field or record part it affects. BCP-05-X-01 therefore turns machine-generated enrichment into a transparent and ignorable annotation rather than an opaque modification of authoritative data.

\subsection{Automatically Enriched Data Streams}

The \texttt{gcve-enriched-dumps} repository illustrates this pattern. It preserves original upstream CVE JSON records while adding AI-assisted severity evaluation, GCVE AI annotation metadata following BCP-05-X-01, and complementary vulnerability-lookup metadata under separate extension fields \cite{gcve-enriched-dumps}. The repository explicitly states that the enriched dumps are not a replacement for original data; they provide context, metadata, AI severity signals, and workflow support for triage and research \cite{gcve-enriched-dumps}.

A generalized enrichment pipeline can be implemented as follows:

\begin{enumerate}
  \item Ingest source records from CVE, GCVE GNAs, or other public vulnerability feeds.
  \item Preserve the original source record unchanged.
  \item Run enrichment tasks such as severity classification, taxonomy tagging, product normalization, relationship discovery, CPE/PURL mapping, or KEV correlation.
  \item Write enrichment results into namespaced extension fields such as \texttt{x\_gcve} and \texttt{vulnerability-lookup:meta}.
  \item Add BCP-05-X-01 provenance metadata describing the AI or automation level, model, review status, and field scope.
  \item Validate records against BCP-05 and extension rules.
  \item Publish dumps as JSON, NDJSON, API output, Git repositories, or static files.
\end{enumerate}

Because unknown extension keys can be ignored by consumers, these enriched streams can evolve rapidly while remaining compatible with conservative parsers \cite{gcve-bcp-05,gcve-bcp-05-x-01,gcve-enriched-dumps}.

\begin{lstlisting}[caption={Illustrative BCP-05-X-01 metadata shape for an enriched record.},label={lst:ai-extension}]
{
  "x_gcve": [
    {
      "extensions": {
        "bcp-05-x-01": {
          "scope": "field",
          "gna_source": 1,
          "field_name": "metrics.ai_severity",
          "ai_level": "augmented",
          "review_status": "none",
          "models": [
            {"name": "VL-AI", "revision": "model-or-commit-id"}
          ],
          "tags": ["severity", "triage", "machine-generated"]
        }
      }
    }
  ]
}
\end{lstlisting}

\section{Distributed KEV Assertions}

Known Exploited Vulnerability information is operationally important because it influences patch prioritization, exposure management, and incident response. Traditional KEV lists, however, are often list-based and opaque. BCP-07 defines KEV as an attributable assertion made by an observer at a time, with status, evidence, scope, timestamps, source, and confidence \cite{gcve-bcp-07}. This separates vulnerability identity from exploitation assertion.

\subsection{Why KEV Must Be Distributed}

Exploit knowledge is not universal. A government catalog, an ISAC, a vendor, a managed detection provider, and a research team may observe different evidence. One actor may classify exploitation as confirmed; another may call it suspected or disputed. BCP-07 therefore allows multiple KEV assertions for the same vulnerability identifier, including CVE, GCVE, and other identifiers \cite{gcve-bcp-07}. The same vulnerability can have multiple statuses over time and across communities, and those differences are not errors; they are evidence of source diversity.

\subsection{BCP-07 Data Model}

A BCP-07 KEV record includes a vulnerability identifier, a KEV status such as confirmed, suspected, disputed, historical, or unknown, and evidence metadata. Additional fields can capture exploitation characteristics, source, confidence, observed dates, scope, and GCVE-specific metadata \cite{gcve-bcp-07}. This makes KEV data suitable for both binary prioritization and richer analytical use.

\begin{lstlisting}[caption={Simplified illustrative KEV assertion model.},label={lst:kev}]
{
  "vulnerability": "GCVE-1-2026-0001",
  "status": "confirmed",
  "evidence": {
    "source": "example-gna",
    "confidence": "high",
    "description": "Observed exploitation in the wild"
  },
  "gcve": {
    "gna": 1,
    "record_type": "kev-assertion"
  }
}
\end{lstlisting}

\subsection{Implementation in vulnerability-lookup}

Vulnerability-lookup 3.0.0 implemented GCVE-BCP-07 support. The release notes state that every vulnerability-lookup instance can publish its own KEV catalog and integrate KEV feeds from CISA and ENISA, using GCVE-BCP-07-compliant conversion and synchronization tooling \cite{vl-3-0-0}. This implementation is significant because it proves that BCP-07 is not only a schema: it can be deployed as a distributed operational model. A local instance can publish a KEV catalog, consume other catalogs, and expose KEV information through user interfaces and APIs.

\section{Reference Implementation: vulnerability-lookup}

Vulnerability-lookup is the reference implementation that makes GCVE practical. GCVE is operated by CIRCL, which also maintains vulnerability-lookup, making the specification and its reference implementation part of the same operational ecosystem \cite{gcve-about}. The vulnerability-lookup user manual notes that the project was designed from the beginning to work independently of vulnerability identifiers, making GCVE integration straightforward, and that it uses the Python GCVE client to retrieve and verify the GCVE registry \cite{vl-gcve-user-manual}.

\subsection{Directory Use and Feed Selection}

Vulnerability-lookup consumes the signed GCVE directory, verifies it, and uses it to discover GNA metadata and publication endpoints \cite{vl-gcve-user-manual,gcve-bcp-01}. It then allows an operator to choose which GNAs or sources to gather from. This embodies the GCVE trust model: the directory is shared, but selection and trust are local.

\subsection{Federated Publication}

Vulnerability-lookup provides the current reference implementation for using the GNA directory, selecting GNAs to gather from, ingesting decentralized publications, and exposing local records through GCVE-compatible APIs. The key architectural point is that decentralized publication is not a private feature of one implementation; BCP-03 defines a reimplementable REST and static-dump publication model, while vulnerability-lookup demonstrates it in production-like software \cite{gcve-bcp-03,vl-home,vl-gcve-user-manual}.

\subsection{Public Database and Correlation}

The January 2026 launch of db.gcve.eu demonstrates vulnerability-lookup as a public open instance that aggregates and correlates vulnerability information from more than 25 public sources, including GCVE GNA sources and established vulnerability databases \cite{gcve-db-announce}. The platform provides a public web interface, API, and open data dumps, showing how GCVE can support both decentralized publication and unified access \cite{gcve-db-announce}.

\subsection{KEV Catalogs}

As described above, vulnerability-lookup 3.0.0 added BCP-07 support, local KEV catalog publication, and integration of CISA and ENISA KEV feeds \cite{vl-3-0-0}. Later releases continued to expose KEV-related views; vulnerability-lookup 5.0.0 includes a \texttt{/kev-catalogs} view listing available KEV catalogs \cite{vl-5-0-0}. The reference implementation therefore turns KEV from a central list into a distributed, source-attributed set of catalogs.

\subsection{CNA/GNA-Compatible Vulnerability Authoring}

Vulnerability-lookup 5.0.0 extended the implementation with a CNA-interoperable API for managing local vulnerabilities and deep Vulnogram integration compatible with CVE 5.2 and GCVE-BCP-05 \cite{vl-5-0-0}. This matters because GCVE adoption depends on authoring workflows, not just ingestion. If GNAs can reserve identifiers, draft records, manage states, and publish through familiar CVD tools, GCVE becomes part of operational disclosure rather than a separate post-processing layer.

\subsection{Support for Enriched Data Products}

The same architecture supports enriched dumps. Vulnerability-lookup metadata can be attached in namespaced fields, AI provenance can be expressed through BCP-05-X-01, and derived dumps can be published without altering original records \cite{gcve-bcp-05-x-01,gcve-enriched-dumps}. This is an example of GCVE's extensibility: data products can be generated automatically while remaining transparent, separable, and ignorable.

\section{Implementation Patterns}

A minimal GCVE-compatible implementation requires only a small number of components:

\begin{enumerate}
  \item a local identifier allocator respecting BCP-04;
  \item a GNA metadata entry in the signed directory;
  \item a BCP-03 publication endpoint or static dump;
  \item BCP-05-compatible records, with explicit relationships when records refer to other records;
  \item BCP-06 conformance metadata describing operational and visibility models;
  \item optional BCP-07 KEV publication, optional BCP-05-X-01 AI provenance, and optional BCP-10 product context.
\end{enumerate}

The following pseudocode captures the pull-side logic used by an aggregator:

\begin{lstlisting}[caption={Simplified GCVE consumer workflow.},label={lst:consumer}]
verify_signature("https://gcve.eu/dist/gcve.json")
directory = load_gna_directory()
trusted_gnas = select_sources(directory, local_policy)

for gna in trusted_gnas:
    endpoint = gna["gcve_pull_api"] or gna["gcve_dump"]
    records = fetch_records(endpoint)
    for record in records:
        validate_bcp05(record)
        index_identifier(record["vulnId"])
        index_relationships(record.get("relationships", []))
        index_extensions(record.get("x_gcve", []))
        store(record, source=gna["id"])
\end{lstlisting}

The publication side is equally simple: allocate a local identifier, produce a BCP-05 record, publish it through the advertised endpoint or dump, and optionally publish related KEV or enrichment objects.

\section{Evaluation and Discussion}

Because GCVE is a socio-technical infrastructure project, evaluation should consider architectural properties rather than only throughput metrics. We evaluate GCVE qualitatively against six properties.

\subsection{Scalability}

GCVE scales allocation by distributing namespace authority to GNAs. It scales publication by allowing direct publication via GNA endpoints or dumps rather than routing every record through a central submission queue \cite{gcve-bcp-03,gcve-bcp-04}. Aggregators such as vulnerability-lookup can scale independently by selecting sources and scheduling pulls according to local needs.

\subsection{Resilience}

A decentralized model reduces dependency on a single publication path. If one GNA endpoint is unavailable, other GNAs remain reachable. If one aggregator disappears, another can consume the same directory and endpoints. The signed directory still plays an important role, but it is smaller and easier to mirror or verify than a central vulnerability database \cite{gcve-bcp-01,gcve-bcp-03}.

\subsection{Semantic Coverage}

BCP-09 substantially increases semantic coverage by allowing GCVE records to represent broader vulnerability-related information objects \cite{gcve-bcp-09}. This improves expressiveness but creates a new requirement: consumers must interpret record types, GNA context, and relationships rather than assuming a uniform canonical description.

\subsection{Trust and Accountability}

GCVE shifts trust from global approval to source transparency and local policy. This is appropriate for a plural vulnerability ecosystem, but it also requires good tooling. Consumers need UI and API support for source provenance, GNA models, signatures, timestamps, review status, AI provenance, and relationship graphs. Vulnerability-lookup is the current reference point for these operational needs \cite{vl-gcve-user-manual,vl-5-0-0}.

\subsection{Extensibility}

BCP-05 and BCP-05-X-01 show that GCVE can evolve through namespaced, ignorable extensions \cite{gcve-bcp-05,gcve-bcp-05-x-01}. The enriched dumps repository demonstrates that automated enrichment can be published in a way that preserves original data and exposes machine-generated provenance \cite{gcve-enriched-dumps}. This design supports fast experimentation without fragmenting the base standard.

\subsection{Operational Adoption}

The January 2026 db.gcve.eu launch and subsequent 2026 vulnerability-lookup releases show adoption through working infrastructure: public database, public API, open dumps, decentralized publication, BCP-07 KEV support, Vulnogram integration, and local vulnerability management workflows \cite{gcve-db-announce,vl-3-0-0,vl-5-0-0}. This is the critical difference between a specification and an operational standard.

\section{Security and Governance Considerations}

GCVE improves autonomy, but autonomy introduces governance challenges.

\subsection{Directory Integrity}

The directory must be verifiable because it maps GNAs to metadata and endpoints. BCP-01 mitigates tampering by requiring signature verification and public key distribution through multiple channels \cite{gcve-bcp-01}. Consumers should fail closed when directory verification fails, log key rotations, and treat endpoint changes as security-relevant events.

\subsection{Publisher Quality and Abuse}

Because GNAs are autonomous, low-quality, misleading, or malicious records are possible. GCVE addresses this through transparency rather than central censorship: GNAs must declare their models, consumers choose whom to trust, and records carry source metadata \cite{gcve-bcp-06}. Aggregators should expose provenance prominently and support allowlists, blocklists, scoring, and conflict display.

\subsection{AI-Generated Data Risk}

AI-assisted enrichment can accelerate triage but can also hallucinate, overstate severity, or introduce systematic bias. BCP-05-X-01 and the enriched dumps design reduce this risk by preserving original data, separating enriched fields, and describing AI level, review status, and model metadata \cite{gcve-bcp-05-x-01,gcve-enriched-dumps}. Consumers should treat unreviewed AI severity as a triage signal, not an authoritative vulnerability assessment.

\subsection{KEV Evidence Quality}

Distributed KEV allows multiple communities to publish exploitation assertions, but it also requires careful evidence handling. BCP-07's status, evidence, source, and confidence fields are essential. UI and automation should avoid collapsing all KEV assertions into a single universal truth unless local policy explicitly defines how to aggregate them \cite{gcve-bcp-07}.

\section{Limitations and Future Work}

GCVE is still evolving. Important future work includes:

\begin{itemize}
  \item conformance test suites for BCP-03, BCP-05, BCP-07, and BCP-05-X-01;
  \item machine-readable GNA conformance dashboards based on BCP-06;
  \item richer relationship ontologies for BCP-09 record graphs;
  \item stronger guidance for conflict representation and duplicate handling;
  \item reproducible enrichment pipelines with signed outputs, model cards, and evaluation data;
  \item broader product and platform graph integration through BCP-10 and cpe-editor;
  \item comparative operational studies of publication latency, coverage, and false-positive rates across centralized, publisher-driven, and decentralized workflows.
\end{itemize}

The most important research challenge is semantic interoperability. GCVE deliberately supports many record meanings. The ecosystem therefore needs tools that can preserve this diversity while still enabling simple operational answers: What affects me? Is it exploited? Which source said so? Is there a patch? Which claim is authoritative for my product or deployment?

\section{Conclusion}

GCVE began as a decentralized identifier allocation system and has become a full-featured vulnerability publication model. Its design combines GNA autonomy, signed discovery, decentralized publication, extensible record formats, source-attributed KEV assertions, AI provenance, enriched data products, and a practical reference implementation in vulnerability-lookup. The most distinctive feature is not the string format of a GCVE identifier; it is the shift from a single canonical vulnerability pipeline to a federated graph of independently governed vulnerability-related information objects.

This shift matches modern vulnerability management. Vulnerabilities are not only discovered; they are interpreted, remediated, contested, exploited, translated, downstreamed, enriched, and operationalized. GCVE provides a standard framework in which those different records can coexist, remain attributable, and be consumed according to local trust policy. Vulnerability-lookup demonstrates that this model is implementable today, while the BCP corpus provides a path for continued evolution without sacrificing interoperability.

\section*{Acknowledgments}

The author thanks the GCVE Working Group, CIRCL, vulnerability-lookup contributors, participating GNAs, and reviewers who have shaped the GCVE specifications and operational tooling.

\printbibliography

\newpage
\begin{appendices}

\section{BCP-to-Implementation Checklist}

\begin{longtable}{p{0.24\linewidth}p{0.68\linewidth}}
\toprule
\textbf{Requirement area} & \textbf{Implementation evidence to collect}\\
\midrule
Directory verification & Signature verification logs, public key source, key rotation process, failure behavior.\\
GNA metadata & Directory entry containing ID, short name, full name, publication URL, API, dump, allocation process, and pull API fields where applicable.\\
Identifier allocation & Local allocator documentation, uniqueness guarantees, non-reuse policy, reserved/rejected states.\\
Publication endpoint & BCP-03 endpoint or static dump, pagination/filter behavior, JSON schema validation, update cadence.\\
Record format & BCP-05 validation, record type handling, relationship fields, extension namespace handling.\\
GNA conformance & Public BCP-06 profile declaring operational model, visibility model, contact, scope, endpoint availability, and update date.\\
KEV catalog & BCP-07 validation, source attribution, status vocabulary, evidence fields, confidence and timestamps.\\
AI extension & BCP-05-X-01 annotations with AI level, review status, field scope, model metadata, and source GNA.\\
Product context & BCP-10 or cpe-editor mappings, UUID stability, aliases, provenance, and moderation workflow.\\
\bottomrule
\end{longtable}

\end{appendices}
\end{document}